\documentclass[aps,prl,reprint,nofootinbib]{revtex4-2}
    \makeatletter
\def\@fnsymbol#1{\ensuremath{\ifcase#1\or *\or ** \or 
    \mathparagraph\or \|\or **\or \dagger\dagger
   \or \ddagger\ddagger \else\@ctrerr\fi}}
    \makeatother
\usepackage{graphicx}
\usepackage{dcolumn}
\usepackage{bm}
	
\usepackage{xcolor}

\usepackage{url}
\usepackage{amsmath} 
\usepackage[T1]{fontenc} 

\usepackage{tikz}
\usepackage{subcaption}

\usepackage{comment}
\usepackage{babel}
\usepackage{esint}
\usepackage[unicode=true,pdfusetitle,
 bookmarks=true,bookmarksnumbered=false,bookmarksopen=false,
 breaklinks=false,pdfborder={0 0 0},pdfborderstyle={},backref=false,colorlinks=false]
 {hyperref}
\hypersetup{
 pdfborderstyle=}

\global\long\def\LL{{\sf \Lambda}}%

\global\long\def\R{\mathbb{R}}%
\global\long\def\C{\mathcal{C}}%

\global\long\def\L{{\Lambda}}%

\global\long\def\d{\mathrm{d}}%

\global\long\def\i{\mathrm{i}}%

\global\long\def\e{\mathrm{e}}%

\newcommand{\Z}{{\mathbb Z}}

\newcommand{\SL}{\mathrm{SL}}

\newcommand{\bea}{{\begin{eqnarray}}}
\newcommand{\eea}{{\end{eqnarray}}}

\makeatletter
\usepackage[parfill]{parskip}
\usepackage{cancel}
\usepackage{babel}

\usepackage{bbold}

\begin{document}
\title{Global Symmetries, Code Ensembles, and Sums Over Geometries}

\author{Ahmed Barbar}
\email{a.barbar@uky.edu}
\author{Anatoly Dymarsky}
\email{a.dymarsky@uky.edu}
\author{Alfred D. Shapere}
\email{shapere@g.uky.edu}
\affiliation{ \\ Department of Physics and Astronomy, \\ University of Kentucky, 506 Library Drive,\\ Lexington, KY, 40506}

\begin{abstract}
We consider Abelian topological quantum field theories (TQFTs) in 3d and show that gaugings of invertible global symmetries naturally give rise to additive codes. These codes emerge as  nonanomalous subgroups of the 1-form symmetry group, parameterizing the fusion rules of condensable TQFT anyons. The boundary theories dual to TQFTs with a  maximal symmetry subgroup gauged, i.e.~with the corresponding anyons condensed,  are ``code'' conformal field theories (CFTs). This observation bridges together, in the holographic sense, results on 1-form symmetries of 3d TQFTs with developments connecting codes to 2d CFTs. Building on this relationship, we proceed to consider the ensemble of maximal gaugings (topological boundary conditions) in a general, not necessarily Abelian 3d TQFT, and propose that the resulting ensemble of boundary CFTs has a holographic description as a gravitational theory -- the bulk TQFT summed over topologies. 
\end{abstract}

\date{\today}

\maketitle
{\it Introduction.} 
Error-correcting codes and related concepts have deeply penetrated many areas of physics, from condensed matter \cite{Kitaev:1997wr,nayak2008non} to field theory \cite{gaiotto_johnson-freyd_2022,Harvey:2020jvu} to quantum gravity \cite{Almheiri:2014lwa,pastawski2015holographic}.   

In the context of quantum field theory, it has been known for several decades that additive codes are related to 2d Abelian CFTs \cite{dolan1996conformal,DOLAN1990165,dong1998framed}, an idea that has gained traction recently \cite{Dymarsky:2020qom}.  The relation between codes and CFTs has led to applications to the modular bootstrap \cite{Dymarsky:2020bps,Dymarsky:2022kwb}, to supersymmetric CFTs \cite{kawabata2023supersymmetric,kawabata2023elliptic}, 
to orbifolds \cite{kawabata2023narain}, and
to holographic ensembles \cite{Henriksson:2022dnu,ADS}.

These advances call for a deeper understanding of
the relation between codes and CFTs, and of its holographic counterpart. In this Letter we address this question and identify how codes emerge in the context of 3d topological quantum field theories that are holographically dual to these 2d CFTs. We find that classical additive codes describe  subgroups of the group of invertible global symmetries of these 3d TQFTs, with even self-dual codes corresponding to maximal nonanomalous (Lagrangian) subgroups. Code-based CFTs emerge as boundary duals to these 3d TQFTs after  nonanomalous symmetries have been gauged. The connection between codes and gaugings of 1-form symmetries opens the door to a new interpretation of many results pertaining to higher-form symmetries \cite{belov2005classification,kapustin2011topological,Gaiotto:2014kfa,hsin2019comments,Buican:2021Galois,Yu2021gauging,benini2023factorization,Shao:2022aik,Buican:2023bzl}, connecting them  with classic problems in information theory and discrete mathematics.

In what follows we first review the relation between codes and 2d Abelian CFTs \cite{Dymarsky:2020qom,angelinos2022optimal}. We then turn to 3d Abelian TQFTs -- specifically Abelian Chern-Simons (CS) theories -- and make an explicit connection between their invertible global symmetries and additive codes. 
We proceed to consider the ensemble of all maximal gaugings of a general 3d TQFT, and propose that it has an equivalent description as a ``gravitational'' theory summing over inequivalent classes of 3d topologies.

{\it Codes and CFTs.}
The code construction of \cite{Dymarsky:2020qom,angelinos2022optimal,ADS} is defined with the help of a lattice $\L\subset \R^{n,\bar{n}}$, which is even with respect to a Lorentzian bilinear form $\eta$ of signature $(n,{\bar n})$, i.e.~the norm squared of any lattice vector defined by $\eta$ is an even integer. In what follows we assume $n={\bar n}$, but our results can be extended easily.
Then the Abelian  group $G=\L^*/\L$ is defined in terms of the dual lattice 
\begin{equation}
\label{dual}
\L^{*}=\{\ell\in\mathbb{R}^{n,\bar{n}}\, |\, \eta(\ell,\ell')\in\mathbb{Z},\,\,\forall\, \ell'\in\L\}.
\end{equation}
It inherits the scalar product $\eta$ from $\L$.
The group $G$ is the  ``dictionary'' -- the full set of all possible ``words.'' A code $\C\subset G$ is a collection of words $c\in G$ closed under addition. A code is even if all its codewords $c\in\C$ are even with respect to $\eta$, $\eta(c,c)\in 2\Z$. 
A code is self-dual with respect to $\eta$ if the only elements of $G$ that have integer scalar product with all the codewords of $\C$ are  themselves the codewords of $\C$.

The crucial step relating codes to lattices is provided by Construction A of Leech and Sloane \cite{leech1971sphere,conway2013sphere}. For any code $\C$ we can define a corresponding lattice 
\begin{equation}
\label{ConstA}
     \Lambda_\C=\{v\,|\, v\in \L^*,\, v\, {\rm mod}\, \L \in \C\},
\end{equation}
which satisfies $\L\subseteq \Lambda_\C\subseteq \L^*$ and $\Lambda_\C/\L=\C$.
When a code is even and/or self-dual, the lattice $\L_\C$ is even and/or self-dual correspondingly. Thus starting from an even self-dual code, from (\ref{ConstA}) we obtain  a so-called Narain lattice, an even self-dual lattice in $\R^{n,{\bar n}}$. A Narain lattice, in turn, defines a Narain CFT, a 2d theory describing e.g.~a string propagating on an $n$-dimensional torus \cite{Polchinski:1998rr}. We call Narain theories defined this way ``code CFTs.''

The group $G$, and the associated codes, remain the same if $\Lambda$ is changed by an $\eta$-preserving $O(n,\bar n, \R)$ transformation, but the resulting Euclidean structure on $\L_\C$ changes. 
The choice of a particular element of $O(n,\bar n, \R)$ in the code CFT construction is known as an embedding \cite{ADS}.

A key property of any code CFT is the representation of its partition function in terms of the underlying code. Consider a CFT based on a code lattice $\L_\C$, on a Riemann surface $\Sigma$ of genus $g$. Its partition function $Z_{\L_\C}$  will be a sum over ``word blocks'' $\Psi_{c_1\dots c_g}$ -- symmetric functions of $g$-tuples of codewords $c_i\in \C$.   Focusing for simplicity on $\Sigma$ of genus one we find 
\begin{equation}
\label{Z}
    Z_\C=W_\C(\{\Psi\}),\quad Z_\C\equiv Z_{\L_\C},\quad  W_\C=\sum_{c\in \C}\Psi_c,
\end{equation}
where $W_\C$ is the so-called code enumerator polynomial \cite{Dymarsky:2020qom}.
The codeword blocks form a finite-dimensional representation of the modular group of $\Sigma$.

It is illustrative to consider the example $\Lambda=\LL \oplus \dots \oplus \LL$, where $\LL$ is a two-dimensional ``glue'' lattice
\begin{eqnarray}
\label{example}
    \LL=\sqrt{k}\left(\begin{array}{cc}
    1 & 0\\
    0 & 1\end{array}
    \right),\quad 
\eta=\bigoplus^n \left(\begin{array}{cc}
    0 & 1\\
    1 & 0\end{array}
    \right).
\end{eqnarray}
In this case a code is a collection of strings of length $2n$ with values in $\Z_k$, closed under addition mod $k$. 
The word blocks $\Psi_c$ form a finite-dimensional representation of  the modular group and $W_\C$ is a polynomial in $k^2$ variables, satisfying certain algebraic identities \cite{angelinos2022optimal,Yahagi:2022idq}. This way of representing the CFT partition function provides a useful way to solve modular bootstrap constraints, essentially reducing them to a classic linear programming problem of Delsarte \cite{delsarte1973algebraic}, and 
leading to a number of recent developments \cite{Dymarsky:2020bps,Dymarsky:2022kwb,Dymarsky:2020pzc,Dymarsky:2021xfc}. 

{\it Abelian TQFTs.}
Next, we review abelian  Chern-Simons theories (CS), and find many parallels with the presentation above. 
$U(1)^{n+\bar{n}}$ CS theory in 3d is defined 
by a lattice $\Lambda\subset\mathbb{R}^{n,\bar n}$ equipped with a bilinear
form, which  here we denote by $K$ (rather than $\eta$), in keeping with the conventional notation in the context of Abelian CS theories \cite{belov2005classification,kapustin2011topological,Cano:2013ooa}.
Large gauge transformations take values in $\Lambda$, which must be an integral lattice
with respect to $K$. We will focus on bosonic CS theories, for which $\Lambda$ must also be even.  Choosing a basis $\{e_{I}\}$ for $\Lambda$ yields 
the CS action on a 3-manifold $M$ in the conventional form 
\begin{equation}
S=\frac{i}{4\pi}\int_{M}K_{IJ}A^{I}\d A^{J},\label{eq:CS action}
\end{equation}
where $K_{IJ}=K(e_{I},e_{J})$ is the K-matrix, the Gram matrix
of $\Lambda$. As above, we focus on $n={\bar n}$ for simplicity.

The Wilson line operators 
\begin{equation}
W_{c}[A;\gamma]=\exp(\i\,c_{I}\oint_{\gamma}A^{I})
\end{equation}
are defined by choosing a closed contour $\gamma$ and a charge $c$, a vector in the dual lattice $\Lambda^{*}$ defined with respect to $K$.
Due to identifications between lines 
\cite{kapustin2011topological}, 
distinct lines  are labeled by the elements
of the discriminant group $c\in G=\Lambda^{*}/\Lambda$.
$G$ is the 1-form symmetry group of the theory \cite{Gaiotto:2014kfa}. It describes the fusion rules of  anyons in the TQFT, with the Wilson lines being anyon worldlines. The anyon spin and braiding statistics are  given by \cite{kaidi2022higher}
\begin{equation}
\theta_{a}=\e^{i\, \pi K^{IJ}a_{I}a_{J}}\label{eq: top_spin},
\end{equation}
\begin{equation}
B_{ab}=\e^{2\pi i\, K^{IJ}a_{I}b_{J}}\label{eq:braiding},
\end{equation}
where $a,b \in G$  are anyon charges. For a Wilson line of charge $a$, $\theta_a$  is the phase acquired by twisting a 
line with framing
by $2\pi$, while $B_{ab}$  is the braiding phase of two lines.

When a CS theory is placed on a handlebody $M$ with boundary $\Sigma$ of genus $g$, the path integral defines a boundary state -- a wavefunction of the CS theory on $\Sigma$. The total dimension of the Hilbert space  of wavefunctions on $\Sigma$ is $\left|G\right|^{g}=\left|\det K\right|^{g}$. 
One can prepare a full basis by inserting Wilson line operators with all possible charges, wrapping all non-contractible cycles of $M$ \cite{witten1989quantum,elitzur1989}.  In the case of genus-one $\Sigma$, when $M$ has only one non-shrinkable cycle $\gamma$, the basis elements  are parameterized by $c\in G$,
\begin{equation}
\label{Psic}
\Psi_{c}[\L]=\int DA\,\,W_{c}[A;\gamma]\,\e^{-S[A]}.
\end{equation}
These are the ``non-holomorphic blocks'' of \cite{Gukov:2004id,belov2005classification}. The path integral on $M$ without Wilson line insertions, i.e., with a trivial Wilson line  of charge $c=0$, gives the partition function $Z_\Lambda$ of a  non-modular-invariant generalized Narain theory based on $\Lambda$ \cite{Narain:1985jj,Ashwinkumar:2021kav},
\bea
\label{holD}
Z_\Lambda({\Sigma})=\Psi_0[\L]\big|_{M}.
\eea

{\it Codes from Symmetries. }
Now we are ready to start bridging together Abelian TQFTs  and  
code CFTs. First, we note that  the 
non-holomorphic blocks (\ref{Psic}) are precisely the word blocks introduced above. In fact,  for the path integral in (\ref{Psic})  to be well-defined, one should specify a compatible Euclidean structure on $\R^{n+\bar n}$.\footnote{The Euclidean structure can be introduced by adding a Maxwell term, which is subsequently decoupled \cite{Gukov:2004id,belov2005classification}. Here we follow the approach of \cite{ADS}, starting with a pure CS theory and specifying the Euclidean structure by a choice of boundary conditions.} Then the possible choices of boundary conditions at $\partial M$ are parameterized 
by $O(n,\bar n, \R)$ modulo equivalences, which we readily recognize as  the possible choices of the embedding parameter of the code CFT construction.


To see how the same code CFT construction emerges from the CS theory, we consider gauging a subgroup of the 1-form symmetry group. In TQFT language this is described as anyon condensation.
In general, 1-form
symmetries can be 't Hooft anomalous \cite{Gaiotto:2014kfa,hsin2019comments}, in which case they cannot be gauged. A subgroup $\C\subset G$
is non-anomalous if the spins and mutual braiding of all of its lines are trivial \cite{kaidi2022higher,benini2023factorization}. 
We readily recognize this as the condition that $\C$ is an even additive code. (In the  case of $\Z_2$ codes this was already pointed out in \cite{Buican:2021uyp} and \cite{Buican:2023bzl}.) 
Gauging the subgroup $\C$ (condensing the anyons of $\C$) is equivalent to summing over insertions of all  Wilson line operators with charges $c\in \C$ \cite{Gaiotto:2014kfa}. 
The resulting theory is a $U(1)^{n+\bar n}$ CS theory based on the lattice
\begin{eqnarray}
\label{gauging}
\Lambda_\C=  \bigcup_{c\in \C}   \Lambda_c,\quad \Lambda_\delta \equiv\{v+\delta\, |\, v\in \Lambda\},
\end{eqnarray} 
which we readily identify with the latice given by Construction A in eq.~(\ref{ConstA}).

A particularly interesting question is to consider  gauging a maximal non-anomalous subgroup $\C$ of $G$. 
In this case the resulting CS theory would be topologically trivial. Indeed, the maximal non-anomalous subgroups are the Lagrangian subgroups, the  subgroups of anyons with trivial mutual braidings, such that all anyons not in the subgroup have nontrivial braiding with at least one anyon inside. This is equivalent to the condition for $\C$, understood as a code, to be even and self-dual; also see \cite{Buicantalk,kawabata2023narain,Buican:2023bzl}.
Accordingly, $\Lambda_\C$ will be a self-dual lattice, with only the trivial line $\Lambda_\C^*/\L_\C=0$ appearing 
in the resulting theory's spectrum. In that theory there is a unique boundary wavefunction given by  (\ref{holD}),
\begin{equation}
\label{hol}
    Z_{\C}=\Psi_0[\L_\C].
\end{equation}
The left side of (\ref{hol}) is the partition function of the code CFT,  the modular-invariant Narain CFT based on an even self-dual $\L_\C$. 
The construction of  
$\Lambda_\C$ from gauging as given by (\ref{gauging}) provides a bulk derivation of (\ref{Z}).

The relation (\ref{hol}) is the holographic duality between the Narain CFT on $\Sigma=\partial M$ and the trivial Chern-Simons theory on $M$.   The boundary conditions for the bulk fields on $\partial M$ are defined in terms of  $\Lambda_\C$; see  \cite{ADS} for the full holographic dictionary.

As is well  known, a Lagrangian subgroup of the 1-form symmetry defines a topological boundary condition in the 3d CS theory \cite{kapustin2011topological,kaidi2022higher}. 
Thus, we find that topological boundary conditions in 3d Abelian Chern-Simons theory are defined by even self-dual codes $\C$.
The state (\ref{hol}) is then given by the 3d path integral on $\Sigma \times [0,1]$ with the topological  boundary condition defined by $\C$ imposed at one of the boundaries. In other words \eqref{hol} is both the boundary state of the  original CS theory on $\Sigma$ and a new, trivial theory obtained after gauging.  

To illustrate the relation between topological boundary conditions and codes, the CS theory defined by (\ref{example}) with $k=2$ is $n$ copies of the low energy limit  of the Toric Code \cite{Kitaev:1997wr,kou2008mutual,Gu:2013gma,Shao:2022aik,Shao:2023gho}. Topological boundary conditions in this theory are defined by real codes of $4^{H_+}$ type \cite{rains2002self}, i.e.,  the even self-dual codes  over $\Z_2\times \Z_2$ that were studied in  detail in \cite{Dymarsky:2020qom}. 

The connection to codes provides an immediate proof that $Z_{\cal C}$ span the space of all modular invariant states of the non-anomalous Abelian CS theory on $\Sigma$, see \cite{nebe2006self}. A simpler version of this result is well-known in the code literature as Gleason's theorem \cite{gleason1971weight,sloane2006gleasonstheoremselfdualcodes}.



Our construction has a natural interpretation  in terms of the SymTFT framework of \cite{Freed:2012bs,Gaiotto:2020iye,apruzzi2023symmetry,kaidi2023symmetry}.
The set of all code CFTs defined by maximal gaugings (\ref{gauging},\ref{hol}) of a given Abelian CS theory forms an orbifold groupoid as defined in \cite{Gaiotto:2020iye}. In particular, the set of code CFTs defined by \eqref{example} 
is the orbifold groupoid of $\Z_k^n$ symmetry. A related picture for $\Z_2$ orbifolds was recently discussed in \cite{kawabata2023narain}.


Up to now we have focused on Narain CFTs and bosonic CS theories. The above picture can be readily generalized to integral but non-even $\L$, connecting gaugings of spin CS theories and Abelian fermionic code CFTs of \cite{Kawabata:2023nlt}. 
It can be also extended to invertible 1-form symmetries of non-Abelian theories. Importantly, codes also emerge  in non-Abelian TQFTs which can condense  into an Abelian phase. In this case additive even self-dual codes of the Abelian phase define topological boundary conditions of the original non-Abelian theory. We give an explicit example of this construction in the End Matter.

One can envision extending the definition of codes to general TQFTs, as mathematical structures parameterizing maximal non-anomalous gaugings and hence topological boundary conditions. While we leave this task for the future, in what follows we will refer to boundary states of an arbitrary TQFT defined by topological boundary conditions as ``code CFTs.''

We emphasize that additive codes of this Letter are {\it different} from  topological (or surface) codes, also appearing in the context of 3d TQFTs \cite{Kitaev:1997wr,nayak2008non,Ellison:2021vth}. The relation between additive codes and topological codes will be addressed elsewhere. 

{\it Holographic ensemble of code CFTs.}
In the remainder of this Letter we would like to apply the concept of averaging over codes, standard in information theory \cite{pless1975classification}, to the holographic setting discussed above. Our starting point is the approach of
\cite{benini2023factorization}, motivated by the expectation that quantum gravity admits  no global symmetries \cite{PhysRevLett.122.191601}.
By gauging a Lagrangian subgroup of the 1-form symmetry of an Abelian CS with a compact gauge group,  \cite{benini2023factorization} arrived at a holographic description of a Narain theory in terms of a trivial bulk TQFT  equivalent to (\ref{hol}).\footnote{Consideration in \cite{benini2023factorization} is limited to left-right factorizable lattices.} Combining this with the ensemble ideas of \cite{Dymarsky:2020pzc,angelinos2022optimal,Kawabata:2022jxt,Henriksson:2022dml,ADS} 
we propose to consider an {\it ensemble} of all maximal gaugings, i.e.~an ensemble of code CFTs using the language of this Letter.

Specifically, starting from any (not necessarily Abelian) 3d TQFT,  we propose to consider an ensemble consisting of all maximal gaugings, and conjecture the resulting ensemble of boundary theories to be dual to the original TQFT coupled to topological gravity, with its partition function given by a sum over all possible topologies. For genus $g=1$ boundary,

\begin{eqnarray}
\label{ensembleH}
    \sum_i \alpha_i\, Z_{\C_i}=\sum_{h \in \Gamma^*\backslash \SL(2,\Z)} h(\Psi_{\rm s}).
\end{eqnarray} 

The sum on the LHS goes over all 
topological boundary conditions 
(even self-dual codes in the Abelian case), labeled by $\C_i$. The   weights of the ensemble $\alpha_i\geq 0$ are determined by the norms of corresponding TQFT states on $\Sigma$ in the limit $g\rightarrow \infty$.
The sum on the RHS is over the mapping class group, interpreted as a sum over equivalence classes of topologies.  In simple cases the  wavefunction $\Psi_{\rm s}$ is the TQFT path integral  on a solid torus.
More generally it is a genus reduction of the TQFT path integral $\Psi_0$ on a handlebody $M$, 
with  $\partial M$ of genus $g\rightarrow \infty$, such that the RHS of \eqref{ensembleH} includes path 
integrals evaluated over non-handlebody topologies. 
Details and a proof of \eqref{ensembleH} 
are given in   \cite{Dymarsky:2024frx}. We give an explicit example evaluating the weights $\alpha_i$ and the seed $\Psi_s$  for a simple non-Abelian TQFT in the End Matter.
  
We note that all examples of AdS${}_{3}$/CFT${}_2$ ensemble holography in the literature readily conform to this framework.
We discuss the  Abelian cases first. The examples of \cite{ADS} correspond to CS theories defined by the lattices  (\ref{example}), and boundary CFTs parametrized by  codes over $\Z_k\times \Z_k$.
The examples of \cite{mukhi2021ads3} correspond to $\Lambda$ being the root lattice of a semi-simple Lie algebra  \cite{angelinos2024complexity,Mizoguchi:2024ahp}.
The examples \cite{raeymaekers2021note,Raeymaekers:2023ras} and 
\cite{Henriksson:2022dml} can also be described by an appropriate $\L$, as will be explained elsewhere. 


The same framework can be applied  to the case of  ``U(1)-gravity'' of \cite{afkhami2021free,maloney2020averaging}, which can be understood, at least in a formal sense, as emerging from a CS theory with  non-compact gauge group $\R^{n+\bar n}$ by gauging all maximal non-anomalous subgroups of the 1-form symmetry group. It is easy to see that in this case  the Lagrangian subgroups are given by even self-dual lattices in $\mathbb{R}^{n,\bar{{n}}}$ \cite{benini2023factorization}, leading to an average over the ensemble of { all} Narain theories on the left-hand side of (\ref{ensembleH}). 

The non-Abelian examples of \cite{Castro:2011zq,mukhi2021poincare,Jian:2019ubz,Romaidis:2023zpx} for a class of $SU(N)_k$ Wess-Zumino-Witten theories  and  minimal models  can also be reformulated as an ensemble of code CFT dual to TQFT gravity, ensuring that only physical theories with positive weights appear in the LHS of \eqref{ensembleH}. We provide an explicit example generalizing \cite{Castro:2011zq,Romaidis:2023zpx,Jian:2019ubz} in the End Matter. 
Likewise, pure 3d quantum gravity of \cite{Maloney:2007ud,Giombi:2008vd,Keller:2014xba}, as formulated recently in terms of Virasoro TQFT \cite{Mikhaylov:2017ngi,collier2023solving,Collier:2024mgv} dual to an ensemble of 2d CFT data \cite{Chandra:2022bqq,Belin:2023efa,deBoer:2024kat}, can be understood within our framework.

{\it Conclusions.} 
In this Letter we have developed a picture connecting the code CFTs of \cite{dolan1996conformal,Dymarsky:2020bps,Yahagi:2022idq,Kawabata:2022jxt,Alam:2023qac} to 3d bosonic Abelian Chern-Simons theories.   Codes arise as subgroups of the 1-form symmetry group; even codes are  non-anomalous subgroups, and even self-dual codes  are  Lagrangian (maximal non-anomalous) subgroups parameterizing topological boundary conditions.
Code CFTs arise on the 2d boundary after gauging these Lagrangian subgroups in the bulk.
This picture readily leads to a representation of  the code CFT partition function (\ref{Z})  in terms of non-holomorphic word blocks \eqref{Psic}.

We have further proposed  that, in the case of a general TQFT, the {\it ensemble of maximal gaugings}, understood as an ensemble of boundary code CFTs, is holographically dual to the original TQFT in the bulk coupled to topological gravity (i.e., summed over topologies). This proposal encompasses all recent examples of ensemble holographic duality \cite{afkhami2021free,maloney2020averaging,Henriksson:2022dml,ADS,mukhi2021poincare,mukhi2021ads3,raeymaekers2021note,Raeymaekers:2023ras,Castro:2011zq,Jian:2019ubz,Romaidis:2023zpx}, and calls for a deeper understanding. One question is whether
(\ref{ensembleH}) admits a generalization to ensembles of non-maximal gaugings, potentially providing an interpretation for the results of \cite{Ashwinkumar:2021kav,Ashwinkumar:2023jtz,Ashwinkumar:2023ctt}. 
Another question is whether ensembles of gaugings, understood as boundary ensembles, can give rise to gravitational theories in a lower or higher number of dimensions.   In $2d$, a natural question would be to understand JT gravity \cite{Saad:2019lba}  in terms of maximal gaugings, given that it can be formulated as a TQFT, namely $\mathrm{PSL}(2,\mathbb{R})$ BF theory \cite{JT2019exact}. Ultimately, it would be interesting to see if the proposed picture based on  gaugings of global symmetries can also include higher-dimensional examples of holography, which so far have eluded an ensemble interpretation.

\begin{acknowledgments}
We thank O.~Aharony, M.~Buican, L.~Eberhardt, Z.~Komargodski, J.~Kulp, and S.~Shao for comments on the manuscript and discussions. AD acknowledges NSF support under grant PHY-2310426.  
\end{acknowledgments}

\bibliographystyle{apsrev4-1}
\bibliography{Refs}

\onecolumngrid 
\vspace{0.5cm}  
\begin{center}
    \textbf{\large End Matter}
\end{center}
\vspace{0.5cm} 
\appendix

\twocolumngrid

Here we provide an explicit example of the holographic correspondence (\ref{ensembleH}) between a {\it non-Abelian} TQFT summed over topologies and a dual ensemble of boundary CFTs $Z_{{\cal C}_i}$ parametrized by codes ${\cal C}_i$.

{\it Appendix A: Topological boundary conditions  of $n$ copies of doubled Ising TQFT via codes}

The Ising TQFT has three anyons denoted as $\{ 1, \epsilon, \sigma \}$. The doubled Ising (DI) TQFT (i.e.~$\text{Ising}\, \times\, \overline{\text{Ising}}$ ) can condense to an intermediate Toric Code (TC) phase by condensing $\epsilon \bar{\epsilon}$  \cite{Bais:2008ni}. In other words, there is an interface (gapped domain wall) between the non-Abelian  DI and the Abelian TC \cite{Bais:2008DW, Lan:2014GDW, Hung:2015GDW}. Starting with $n$ copies of the doubled Ising TQFT, one can achieve a maximal (Lagrangian) condensation by first condensing to an intermediate phase of $n$ TCs, and then condensing a maximal (Lagrangian) subgroup of the Abelian 1-form symmetry group parametrized by a code $\cal C$, as explained in  the main text. Equivalently, topological boundary conditions for $n$ copies of the TC, parameterized by additive codes $\cal C$ of real $4^{H+}$ type,  define topological boundary conditions of the original $n$ copies of the DI on the other side of the interface.
This parametrization is related to the $C$ and $D$ codes of  \cite{dong1998framed}, as will be discussed elsewhere. For $n<8$ all  maximal condensations can be obtained  this way. 

The TC can be described as a Chern-Simons theory defined by the lattice \eqref{example} with $k=2, n=1$. The anyons  $1$, $e$, $m$ and $f$ of the TC  are Wilson lines with charges 
$c \in G=\Lambda^*/\Lambda=\mathbb{Z}_2 \times \mathbb{Z}_2$, 
namely $(0,0)$, $(0,1)$, $(1,0)$ and $(1,1)$ respectively. Electromagnetic duality of the TC is a surface operator $E$ exchanging $e$ and $m$. It is a $\Z_2$ subgroup  of the $O(1,1,\Z)$ symmetry of the bulk theory \cite{ADS}. At the level of charges   $c\in G$  it acts by mapping $c=(\alpha,\beta)$ to $c'=(\beta,\alpha)$.
The interface between the DI and the TC, placed on a toroidal cylinder, maps the wavefunctions \eqref{Psic} of the latter to the wavefunctions of the former, 
\begin{equation}
\label{gen_1_TC}
\begin{aligned}\Psi_{00}  =\left|\chi_{1}\right|^{2}+\left|\chi_{\epsilon}\right|^{2},\quad 
\Psi_{01}  =\left|\chi_{\sigma}\right|^{2},\\
\Psi_{10}  =\left|\chi_{\sigma}\right|^{2},\quad 
\Psi_{11}  =\chi_{1}\bar{\chi}_{\epsilon}+\chi_{\epsilon}\bar{\chi}_{1}.
\end{aligned}
\end{equation}
Similar relations hold for any Riemann surface $\Sigma$. 

There are two even self-dual codes of length $n=1$ with the scalar product \eqref{example} and $k=2$, ${\cal C}=\{(0,0),(1,0)\}$ and ${\cal C}'=\{(0,0),(0,1)\}$, related by electromagnetic duality (code equivalence) \cite{Dymarsky:2020bps}. Corresponding maximal gaugings define the boundary state $|Z_{\cal C}\rangle=|Z_{{\cal C}'}\rangle$ of the DI theory on a Riemann surface $\Sigma$. In case when $\Sigma$ is a torus, 
\begin{equation}
    Z_{{\cal C}}=Z_{{\cal C}'}=Z_{\text{Ising}}=\left|\chi_{1}\right|^{2}+\left|\chi_{\epsilon}\right|^{2}+\left|\chi_{\sigma}\right|^{2}.
\end{equation}


Now let us consider our main example, $n=2$ copies of the DI TQFT. For convenience we will refer to it as $\text{DI}^2$ theory. 
There is an interface between $\text{DI}^2$ and two copies of the TC, which we will denote as TC${}^2$. There are 6 distinct real $4^{H+}$ codes of length $n=2$ parametrizing  topological boundary conditions of the latter. The  codes split into two classes under the  $\Z_2 \times \Z_2$ ``electromagnetic'' group of code equivalences \cite{Dymarsky:2020qom}, giving rise to two maximal gaugings  of $\text{DI}^2$ theory. To be explicit, we write a code representative from each of the two equivalence classes,
\begin{equation}
\label{codesn2}
\begin{aligned}\mathcal{C}_{1} & =\left\{ (00,00),(00,01),(01,00),(01,01)\right\}, \\
\mathcal{C}_{2} & =\left\{ (00,00),(01,01),(10,10),(11,11)\right\}.
\end{aligned}
\end{equation}
Corresponding boundary states on the torus -- the torus partition functions  of  the Ising-squared  CFT  and its $\Z_2$ orbifold, the compact boson of radius $R=2$ (bosonized Dirac fermion) -- are as follows
\begin{equation}
\begin{aligned}Z_{\text{Ising}^2}\equiv Z_{1} & =\left(\left|\chi_{1}\right|^{2}+\left|\chi_{\epsilon}\right|^{2}+\left|\chi_{\sigma}\right|^{2}\right)^{2},\\
Z_{\text{Dirac}}\equiv Z_{2} & =\left|\chi_{1}^{2}+\chi_{\epsilon}^{2}\right|^{2}+4\left|\chi_{1}\chi_{\epsilon}\right|^{2}+2\left|\chi_{\sigma}^{2}\right|^{2}.
\end{aligned}
\end{equation}
There are, in fact, two distinct ways in which two copies of the DI TQFT can be interfaced with (condense to) two copies of the TC. The two choices are related by the interchange of the two Isings while the two copies of  $\overline{\text{Ising}}$ remain intact. In this way one can obtain two more topological boundary condition of the $\text{DI}^2$ theory. It turns out that in both cases the code ${\cal C}_2$ defines the same topological boundary condition, but the topological boundary conditions defined by ${\cal C}_1$ are different. We  denote the corresponding states in the Hilbert space of $\text{DI}^2$ theory on a Riemann surface $\Sigma$ by $|Z_1\rangle$ and $|Z_1'\rangle$. We note that as CFTs, $Z_1$ and $Z_1'$ are the same, as they  only differ by a relabeling of the characters $\chi^1_a$ and $\chi^2_a$, $a\in\{1,\epsilon,\sigma\}$.

{\it Appendix B: Weights of the holographic ensemble}

\noindent The  states $|Z_1\rangle,|Z_1'\rangle,|Z_2\rangle$ of the $\text{DI}^2$ theory on $\Sigma$ defined by the topological boundary conditions of this theory constitute the boundary ensemble holographically dual, in the sense of \eqref{ensembleH}, to $\text{DI}^2$ theory summed over all possible topologies. Below we evaluate the relative weights $\alpha_i$, using the result of   \cite{Dymarsky:2024frx} for  their ratio 
\begin{equation}
\label{weights}
\frac{\alpha_{i}}{\alpha_{j}}=\lim_{g\rightarrow\infty}\frac{\left\langle Z_{j}|Z_{j}\right\rangle }{\left\langle Z_{i}|Z_{i}\right\rangle }.
\end{equation}
The task here is to calculate the norms of the corresponding  states on a $\Sigma$ of arbitrary large genus $g$. For an Abelian TQFT this would be simple because the Hilbert space on $\Sigma$ of genus $g$ is  just a tensor product of $g$ Hilbert spaces for the torus, and the norm of a state $|Z_{\cal C}\rangle$, specified by the topological boundary condition parametrized by the code $\mathcal{C}$,  has the norm 
\begin{equation}
\label{Znorm}
\left\langle Z_{\mathcal{C}}|Z_{\mathcal{C}}\right\rangle =\left|\mathcal{C}\right|^{g}.
\end{equation}

For a non-Abelian TQFT such a calculation would be nontrivial. Below we leverage the  Abelian intermediate phase of the $\text{DI}^2$ theory to develop a computational technique evaluating \eqref{weights}.
The main idea is as follows. Any state $|Z_i\rangle$  in $\text{DI}^2$  theory can be defined through an interface by the topological boundary conditions in the Abelian TC${}^2$ theory. Hence the norm $\langle Z_i|Z_i\rangle$ can be defined by an appropriate surface operator $\mathcal{G}$ in the Abelian TC${}^2$ theory, 
\begin{equation}
\left\langle Z_{i}|Z_{i}\right\rangle =\left\langle Z_{\mathcal{C}_{i}}\right|\mathcal{G}\left|Z_{\mathcal{C}_{i}}\right\rangle.
\end{equation}
We illustrate this in Fig.~\ref{fig_interface}.

\begin{figure}
\tikzset{every picture/.style={line width=0.75pt}} 

\begin{tikzpicture}[x=0.75pt,y=0.75pt,yscale=-1,xscale=1]

\draw   (190.8,129.76) -- (329.47,129.76) -- (329.47,170.7) -- (190.8,170.7) -- cycle ;
\draw [color={rgb, 255:red, 208; green, 2; blue, 27 }  ,draw opacity=1 ][line width=0.75]    (240.32,130.16) -- (240.3,170.7) ;
\draw [color={rgb, 255:red, 208; green, 2; blue, 27 }  ,draw opacity=1 ][line width=0.75]    (280.59,130.09) -- (280.55,171.2) ;
\draw   (380.5,130.68) -- (480.2,130.68) -- (480.2,170.6) -- (380.5,170.6) -- cycle ;
\draw [color={rgb, 255:red, 208; green, 2; blue, 27 }  ,draw opacity=1 ][line width=1.5]    (430.25,130.59) -- (430.25,170.85) ;
\draw    (339.75,149.7) -- (367.75,149.93) ;
\draw [shift={(369.75,149.95)}, rotate = 180.48] [color={rgb, 255:red, 0; green, 0; blue, 0 }  ][line width=0.75]    (10.93,-3.29) .. controls (6.95,-1.4) and (3.31,-0.3) .. (0,0) .. controls (3.31,0.3) and (6.95,1.4) .. (10.93,3.29)   ;

\draw (202.4,140.8) node [anchor=north west][inner sep=0.75pt]  [font=\large] [align=left] {TC};
\draw (291.73,141.07) node [anchor=north west][inner sep=0.75pt]  [font=\large] [align=left] {TC};
\draw (250.73,141.07) node [anchor=north west][inner sep=0.75pt]  [font=\large] [align=left] {DI};
\draw (392.23,141.72) node [anchor=north west][inner sep=0.75pt]  [font=\large] [align=left] {TC};
\draw (441.57,141.98) node [anchor=north west][inner sep=0.75pt]  [font=\large] [align=left] {TC};
\draw (424.8,174.9) node [anchor=north west][inner sep=0.75pt]  [font=\footnotesize,color={rgb, 255:red, 208; green, 2; blue, 27 }  ,opacity=1 ] [align=left] {$\displaystyle \mathcal{G}$};

\end{tikzpicture}

  \caption{Visual representation of the surface operator $\mathcal{G}$.}
  \label{fig_interface}
\end{figure}

For the doubled Ising condensing into the TC, this surface operator $\mathcal{G}$ is simply the projector on the sector invariant under  ``electromagnetic'' symmetry,
\begin{equation}
\label{n=1}
\mathcal{G}=2^{g-1}\left({I+E}\right),
\end{equation}
where $I$ is the identity and $E$ is the surface operator that exchanges $e \leftrightarrow m$ anyons of the TC theory.
To illustrate the utility of \eqref{n=1} we calculate the norm of the only invariant of the doubled Ising theory -- the Ising model, 
\begin{equation}
\label{Ising_norm_TC}
\begin{split}\left\langle Z_{\text{Ising}}|Z_{\text{Ising}}\right\rangle  & =2^{g-1}\left(\left\langle Z_{\{00,01\}}|Z_{\{00,01\}}\right\rangle +\right.\\
&\qquad \qquad\left.
\left\langle Z_{\{00,01\}}|Z_{\{00,10\}}\right\rangle \right)  \\
 &=2^{g-1}\left(2^{g}+1\right),
\end{split}
\end{equation}
where we used a generalization of \eqref{Znorm}, $\left\langle Z_{\mathcal{C}_{i}}|Z_{\mathcal{C}_{j}}\right\rangle =\left|\mathcal{C}_{i}\cap\mathcal{C}_{j}\right|^{g}$.
One can confirm \eqref{Ising_norm_TC} using the fact that $Z_{\text{Ising}}$ is a diagonal invariant, and so its norm is the same as the Hilbert space dimension of the  (chiral) Ising TQFT placed on $\Sigma$. The latter is given by \cite{Barkeshli2019}
\begin{equation}
\mathrm{dim}\,\mathcal{H}_{g}=\mathcal{D}^{2g-2}\sum_{a}d_{a}^{2-2g},
\end{equation}
where $\mathcal{D}$ is the total quantum dimension and $d_a$ are the quantum dimensions of the anyons.
Applying this to the Ising theory gives
\begin{eqnarray}
\left\langle Z_{\text{Ising}}|Z_{\text{Ising}}\right\rangle  & =&\mathrm{dim}\, \mathcal{H}_{g}^{\text{Ising}}\\
 & =&2^{2g-2}\left(1+1+(\sqrt{2})^{2-2g}\right), 
\end{eqnarray}
in agreement with (\ref{Ising_norm_TC}).

Now we extend this technique to the case of  $\text{DI}^2$ theory. The surface operator is simply the tensor product of two $\mathcal{G}$ such that 
\begin{equation}
\label{Z2_norm_eq}
\begin{split}
&\left\langle Z_{i}|Z_{\text{i}}\right\rangle  =\left\langle Z_{\mathcal{C}_{i}}\right|\mathcal{G}\otimes\mathcal{G}\left|Z_{\mathcal{C}_{i}}\right\rangle \\ \nonumber
 & =2^{2g-2}\left\langle Z_{\mathcal{C}_{i}}\right|\left(I\otimes I+I\otimes E+E\otimes I+E\otimes E\right)\left|Z_{\mathcal{C}_{i}}\right\rangle. 
\end{split}
\end{equation}
Applying this to the codes in \eqref{codesn2} yields 
\begin{eqnarray}
\left\langle Z^{}_{1}|Z^{}_{\text{1}}\right\rangle =\langle Z_{1}^{'}|Z_{\text{1}}^{'}\rangle &=&2^{2g-2}\left(2^{g}+1\right)^{2},\\
\left\langle Z_{2}|Z_{\text{2}}\right\rangle &=&2^{3g-1}\left(2^{g}+1\right).
\end{eqnarray}

Now using (\ref{weights}) we readily find $\alpha_{1}=\alpha_{1}^{'}=2\alpha_{2}$.
Normalizing the weights such that they add up to unity, and taking into account that $|Z_1\rangle$ and $|Z_1'\rangle$ define the same CFT, we find the LHS of \eqref{ensembleH} to be 
\begin{equation}
\label{holex}
\langle Z\rangle_{\rm boundary}=\frac{4}{5}Z_{\text{Ising}^{2}}+\frac{1}{5}Z_{\text{Dirac}}.
\end{equation}
\noindent {\it Appendix C: Sum over topologies}

\noindent The holographic duality proposed in the main text is the equality between the averaged CFT partition function on $\Sigma$ of genus $g$ and the bulk TQFT summed over all 3d topologies. This sum is defined as the sum of the vacuum character over the mapping class group of a Riemann surface of genus $g' \rightarrow \infty$, with subsequent genus reduction to finite $g$ \cite{Dymarsky:2024frx}.  
Equation \eqref{ensembleH} manifests this duality in the simple case of $g=1$ when $\Sigma$ is a torus. To evaluate the seed wavefunction $\Psi_s$ one should start with the TQFT path integral on a handlebody ending on  $\Sigma'$, reduce to $g=1$, and then take the limit $g' \rightarrow \infty$. The result should be averaged over {\it all} handlebodies ending on $\Sigma'$. For a non-Abelian theory this is a complicated task, but the partial condensation to an Abelian phase helps here as well. For the DI TQFT, the seed on the torus is, up to a coefficient, simply the seed for the TC which is its vacuum character \cite{ADS}. Thus, from \eqref{gen_1_TC} we find $\Psi_s={1\over 2}(\left|\chi_{1}\right|^{2}+\left|\chi_{\epsilon}\right|^{2})$, 
\begin{eqnarray}
    Z_{\rm Ising}(\tau)=\sum_{h\in \Gamma_0(2)\backslash SL(2,\Z)} \Psi_s(h\, \tau).
\end{eqnarray}
For the $\text{DI}^2$ theory the situation is more complicated because there are two ways to condense this theory to TC${}^2$, with the state $|Z_2\rangle$ present in both cases. The seed, therefore, is a sum of images of  the TC${}^2$ vacuum character $\Psi_{00}^2$ mapped under \eqref{gen_1_TC}, with the overlapping state $Z_2$ subtracted (a detailed derivation of this result will be given elsewhere), 
\begin{eqnarray}
\nonumber
    \Psi_s=\frac{4}{5}(\left|\chi_{1}\right|^{2}+\left|\chi_{\epsilon}\right|^{2})^2-\frac{1}{15} Z_{\rm Dirac}. 
\end{eqnarray}
One can check straightforwardly that the ``Poincare series'' of this seed matches \eqref{holex},
\begin{eqnarray}
    \langle Z(\tau)\rangle =\sum_{h\in \Gamma_0(2)\backslash SL(2,\Z)} \Psi_s(h\, \tau).
\end{eqnarray}

\end{document}